\documentclass[prl,preprintnumbers,twocolumn,aps,superscriptaddress]{revtex4}
\usepackage{graphicx,}
\usepackage{graphicx,dcolumn,bm,epsfig,amsmath,amssymb,}
\usepackage{float}

\begin{document}

\title {Pulsars as Weber gravitational wave detectors}

\author{Arpan Das}
\email{arpan@iopb.res.in}
\affiliation{Institute of Physics, Bhubaneswar 750115, India}
\affiliation{Homi Bhabha National Institute, Training School Complex,
Anushakti Nagar, Mumbai 400085, India}

\author{Shreyansh S. Dave}
\email{shreyansh@iopb.res.in}
\affiliation{Institute of Physics, Bhubaneswar 750115, India}
\affiliation{Homi Bhabha National Institute, Training School Complex,
Anushakti Nagar, Mumbai 400085, India}

\author{Oindrila Ganguly}
\email{oindrila@iopb.res.in}              
\affiliation{Institute of Physics, Bhubaneswar 750115, India}

\author{Ajit M. Srivastava}
\email{ajit@iopb.res.in}
\affiliation{Institute of Physics, Bhubaneswar 750115, India}
\affiliation{Homi Bhabha National Institute, Training School Complex,
Anushakti Nagar, Mumbai 400085, India}


\begin{abstract}
A gravitational wave passing through a pulsar will lead to a
variation in the moment of inertia of the pulsar affecting its rotation.
This will affect the extremely accurately measured spin rate of the
pulsar as well as its pulse profile (due to 
induced wobbling depending on the source direction). The effect will be 
most pronounced at resonance and should be detectable by accurate 
observations of the pulsar signal. The pulsar, in this sense, acts as a 
remotely stationed Weber detector of gravitational waves whose signal can 
be monitored on earth. With possible gravitational wave sources spread 
around in the universe, pulsars in their neighborhoods can provide us a 
family of \textit{remote} detectors all of which can be monitored on  earth.
Even if GW are detected directly by earth based conventional detectors, such 
pulsar detectors can provide additional information for accurate 
determination of the source location. This can be of crucial importance for 
sources which do not emit any other form of radiation such as black hole 
mergers. For the gravitational wave events already detected by LIGO 
(and VIRGO), our proposal suggests that one should look for specific pulsars
which would have been disturbed by these events, and will transmit this
disturbance via their pulse signals in any foreseeable future. If these
future pulsar events can be predicted with accuracy then a focused effort 
can be made to detect any possible changes in the signals of those specific
pulsars.
\end{abstract}

\maketitle

\noindent 

Observation of gravitational waves (GW) by LIGO \cite{abbott2016} has opened 
a new window to the universe where even events of black hole mergers can be 
detected which otherwise leave no signatures in the electromagnetic spectrum. 
The most recent observation of binary neutron star (BNS) merger via 
gravitational waves, along with electromagnetic radiation, has provided us 
the opportunity to probe such events with clear identification of the source 
\cite{abbott2017}. As more and more gravitational wave detectors are set up 
around the globe, our ability to detect gravitational waves with good
localization of the source in the sky will improve tremendously. Future  
space-based detectors will further complement the search for gravitational 
wave sources with wide range of wavelengths and strengths.
The ultimate limitation on these earth based and space-based detectors
will arise from two main factors. Our location is one such factor
as most of the powerful sources of gravitational waves are likely to occur 
very far. Secondly, the ability of our near-earth detectors in triangulating
the location of the sources will be limited by the very nature
of the detectors. One would wish if a family of detectors could be placed 
far away in space, and then signals from these \textit{far away} detectors 
could be collected with high precision for GW detection and accurate 
determination of the location of the source (possibly in conjunction
with conventional near earth detectors). We propose such a possibility 
in this paper. 

Main physics underlying our proposal is based on the fact 
that with accurate measurements of timings of pulsars (rapidly rotating
neutron stars) very minute changes in the moment of inertia (MI) of the 
neutron star (NS) may be observable
providing a sensitive probe for its configurational changes.
This should act as a detector of a gravitational wave passing through 
the  NS which will lead to transient deformations of the NS 
changing its moment of inertia. Depending on the angle between the GW 
source direction and the pulsar rotation axis, wobbling of the pulsar 
may also be induced. Hence, a combination of detection of
changes in the rotation rate, along with the changes in the pulse profile, 
will allow probing the strength of the strain amplitude from
GW as well as its direction of propagation w.r.t the NS rotation
axis. Clearly, the effects will be most pronounced at the resonance
when the GW frequency matches a natural frequency of the NS.
The neutron star, in this sense, acts like a Weber gravitational wave
detector \cite{weber,aguiar2011,ring}, with the GW induced oscillations 
of the star, which get imprinted on its pulses, being detected via 
variations in its rotation pattern. Thousands of pulsars 
(most within our galaxy), which are accurately monitored, thus act as a 
family of \textit{remotely stationed} Weber detectors of gravitational waves.
(Or, for that matter, for any astrophysical event which disturbs the 
moment of inertia of the pulsar.) This can extend our reach 
of the detection of GW sources further out in space. At present, as 
most of the observed pulsars are within our galaxy, GW sources
detectable by them are likely to be nearby. These are then very likely to be
directly detected by earth/space based conventional detectors. Even then,
such pulsar detectors, especially when they lie near that source, 
will be able to provide additional information
about the source direction, improving the accuracy of  determination of 
the source location. This is of crucial importance for sources which do 
not emit any other form of radiation such as black hole mergers.
For example, the recent BNS merger detection by LIGO was supplemented
with gamma ray and optical observations for precise identification of the 
source. This could not be possible for the first LIGO detection of GW 
originating from merger of black holes where except GW, no other forms 
of radiation was supposedly emitted. Even with a future large network of 
earth/space based conventional GW detectors, it is unlikely to achieve 
high enough angular resolution for precise determination of the location 
of such  distant pure GW-sources. In such a case, any other remotely stationed 
detector (such as these pulsar detectors), can provide additional crucial 
information. Our proposal also provides an opportunity to, in some sense,
revisit the  gravitational wave events which have already been detected by LIGO 
(and VIRGO). For this one should look for specific pulsars
which would have been disturbed by these events, and will transmit this
disturbance via their pulse signals in any foreseeable future. If these
future pulsar events can be predicted with accuracy then a focused effort 
can be made to detect any possible changes in the signals of those specific
pulsars. This will be a direct test of any feasibility of the present proposal.
Any possible such detection will give valuable information about
the neutron star parameters (such as equation of state).

We mention that our proposal is fundamentally different from the conventional 
method of using pulsar timing array for the detection of gravitational
waves \cite{pta}. In the latter case, pulse arrival times are monitored for
gradual changes due to passage of very long wavelength gravitational waves 
in the intervening region. Thus, it is limited to extremely low 
frequency sources with typical frequencies less than $10^{-6}$ Hz arising 
from rare sources such as supermassive black hole mergers or exotic 
proposed objects such as cosmic strings. Further, with pulsar timing array,
to get any information about the direction of GW source(s) a network of 
pulsars will be required. In contrast, in principle, a single pulsar 
detector as proposed here, through changes in its spin rate and pulse 
profile, can give some information about the GW source direction.

A gravitational wave represents periodic variations in the metric of
spacetime. Associated  changes in the Riemann curvature tensor 
$R_{\mu\nu\lambda\rho}$ lead to deformations of a matter body
 through which
the GW is passing. For a neutron star, changes in its quadrupole moment
$Q_{ij}$ resulting from a GW can be written in the following 
form \cite{hinderer2008}:
\begin{equation}
Q_{ij} = - \lambda_d E_{ij}~,
\label{eq:qij}
\end{equation}
where $E_{ij}$ is the external tidal field and $\lambda_d$ is the tidal deformability given by 
\begin{equation}
\lambda_d = \frac{2}{3} k_2 \frac{R^5}{G}.  
\label{eq:lambda}
\end{equation}
Here, $k_2$ is the second Love number with its value for a NS lying in the 
range $k_2 \simeq 0.05 - 0.15$ \cite{abbott2017} and $R$ denotes the radius 
of the neutron star. Eqn.(1) is valid for static gravitational fields, hence
applicable only for wavelengths larger than the NS radius $R$. This will
hold true for the
range of frequencies we are interested in (kHz or smaller, these being
the typical frequencies for BNS merger and also where NS resonant
modes are expected).

 We can write $E_{ij} = R_{i0j0}$ in terms of the GW 
strain amplitude $h$ for a specific polarization. In transverse traceless 
(TT) gauge,
\begin{equation}
R_{\mu 0 \nu 0} = - \frac{1}{2}\partial_0\partial_0h_{\mu\nu}^{TT}
\label{eq:ttriemann}
\end{equation}
where
\begin{equation}
h_{\mu\nu}^{TT} = C_{\mu\nu}e^{ik_{\sigma}x^{\sigma}}
\label{eq:h}
\end{equation}
and
\begin{equation}
C_{\mu\nu}=
\begin{pmatrix} 
0 & 0 & 0 & 0 \\
0 & h_{+} & h_{\times} & 0\\
0 & h_{\times} & -h_{+} & 0\\
0 & 0 & 0 & 0
\end{pmatrix}
\label{eq:c}
\end{equation}
The suffixes `+' and `$\times$' denote two different polarizations of GW. Thus, for the $+$ polarization, the 
amplitude of resulting $E_{ij}$ is given by (in the transverse-traceless 
gauge) \cite{carrollbook}:
\begin{equation}
E_{xx} = - E_{yy} = \frac{2\pi^2 h c^2}{\lambda^2},
\label{eq:ett}
\end{equation}
$\lambda$ being the wavelength of GW. Here we use $h$ to denote $h_+$. 
Resulting changes in the quadrupole moment tensor are given by 
eq. \eqref{eq:qij}. For simplicity, we take 
the initial configuration of NS to be spherically symmetric and the
deformation to be ellipsoidal (with the dimension in the
direction of propagation of GW remaining unchanged).
Using values $Q_{ij}$ obtained from eq. \eqref{eq:qij}, we can then  
calculate changes in the moment of inertia tensor $I_{ij}$ of the neutron 
star as follows (assuming ellipsoidal deformation of an initially spherical
object)

\begin{equation}
\frac{\Delta I_{xx}}{I} = - \frac{\Delta I_{yy}}{I}
\simeq \frac{k_2}{3} \frac{R^3 c^2}{GM \lambda^2} 20h
\label{eq:deltami}
\end{equation}
where $M$ is the mass of the neutron star. We take sample values with 
$M = 1.0 M_\odot$ and $R = 10$ km. Highest sensitivity will be
reached for smallest values of $\lambda$.  (Still retaining the validity 
of Eqn.(1) requiring $\lambda$ much larger than NS radius $R$, as we will
see range of frequencies we consider are below kHz). As a typical astrophysical
source of GW, we can take binary NS merger (such as the one recently 
observed by LIGO and Virgo) with the highest value of GW frequency being 
about 1 kHz 
Then, we get (taking $k_2$ = 0.1 as a sample value)  
\begin{equation}
\frac{\Delta I_{xx}}{I} \simeq 10^{-2} h~.
\label{eq:sampledeltami}
\end{equation}

The BNS merger recently detected by LIGO-Virgo had
peak strength of the signal $h \simeq 10^{-19}$ \cite{abbott2017}.  
Source distance has been estimated to be about 130 million light years.
We can then imagine that our prototype pulsar detector is in the vicinity
of such a GW source with an optimistic separation between the source and 
the pulsar detector being about one light year (see, footnote 
\footnote{We note that this is not 
likely a realistic possibility with the present stage of observations of 
pulsars as most observed pulsars are within our galaxy, the most distant 
being about 50 Mly away. Thus, these particular numbers may be taken
as a future possibility. We may say that our suggestion should provide 
motivation for detailed and precision studies of pulsars very far away.
For nearby sources, the strength of this technique lies in its ability
to give additional information about the GW source direction as mentioned 
above.}).
Resulting value of $h$ at the pulsar location will then be enhanced to 
$h \simeq 10^{-11}$. Fractional changes in the spin rate $\nu$ of the 
pulsar will be the same as the fractional change in the MI. Thus, we 
estimate changes in the spin rate of the pulsar to be
\begin{equation}
\frac{\Delta \nu}{\nu} = \frac{\Delta I}{I} \simeq 10^{-13}~.
\end{equation}
Here, we have written $\Delta I$ for the change in the relevant component
of MI.  Changes in the fractional spin rate of the pulsar of this
order should be detectable by precision measurements of the pulses.
This is certainly true for certain millisecond pulsars whose pulse timings 
have been measured to an accuracy of about $10^{-15}$ seconds (or 
better, thereby increasing the sensitivity of these pulsar detectors). 
One may be concerned whether this accuracy will apply to the changes 
induced by the GW pulse as typically one needs to fold a very large number 
of pulses to get high accuracy. For this issue it is important to note 
that the neutron star acting as a Weber detector at resonance will exhibit
{\it ringing} effect. Thus, even for a GW pulse, its effect will be 
present in the pulsar signal for a much longer duration.

Clearly, with higher precision measurements of pulsar timings,
weaker, or more distant sources (from the detector-pulsar) may be
detected, increasing the frequency of such detections. We have discussed 
BNS merger as one particular example of a source. Gravitational waves from
any other source, with suitable amplitude may be detected in this
manner. Note that the duration of the changes in the pulse 
amplitude (and its profile) depends on the source and can be large,
varying from a millisecond to hundreds of seconds (as for the recent BNS 
merger). The pulsar timings will then show variations which may be a sharp
variation, or a slow modulation extending over many pulse periods
(depending on the pulsar spin rate). Further note that for a generic direction
of propagation of the GW, it will induce wobbling of the pulsar (on top of any
already present). This will lead to a modulation of the pulse intensity
profile. Detailed simulations of these changes can give information about
relative direction of GW propagation w.r.t the rotation axis of NS. 
Such signals could be searched for in the recorded data of pulsar timings,
looking for sudden transient changes, as well as relatively longer modulations
of the pulses. It may be noted that the GW induced sudden changes should in
general be different from the standard glitches where recovery time scales
are much longer. We also note that GW induced transient change in the spin 
rate may be positive as well as negative, and so could also appear as
an anti-glitch \cite{archibald2013} (which cannot be explained within the 
framework of the standard vortex depinning model of glitches).

Note that the above estimates of changes of spin rate of NS have not
accounted for resonances which can dramatically increase the effects of
GW. (For example, resonant tidal deformations from orbiting BNS leading 
to rupture of NS crust have been discussed in the literature 
\cite{tsang2012}. For specific modes, the resonant frequencies
for NS can be in the range of 100 Hz to 1 kHz, which will be relevant for a 
typical BNS merger GW source.) Under the influence of a gravitational
wave the amplitude for a specific mode with natural 
frequency $\omega$ can be written as \cite{nstaroscl},

\begin{equation}
\xi \propto {1 \over \omega^2 - 4\Delta\Omega^2 -
2i\Delta\Omega/t_{visc}} e^{2i\Delta\Omega t}
\end{equation}
 
 where $\Delta\Omega = \Omega - \Omega_s$, $\Omega$ being
 the driving (GW) frequency and $\Omega_s$ the spin
 angular frequency of the pulsar. $t_{visc}$ is the 
 time-scale for dissipation. It has been argued in \cite{nstaroscl}
 that $t_{visc}$ is much larger than the orbital decay time scale
 $t_D$ of the binary neutron star system (even much before merger). 
 As the time scale for GW  pulse we are considering will be much 
 smaller, it is possible that $t_{visc}$ can be assumed to be large 
 here as well, thereby leading to significant enhancement of the 
 amplitude at resonance when $\omega = 2\Delta \Omega$. 
 (For certain alignment between the spin axis and 
 the direction of the gravitational wave, there may be
 an averaging effect, which will suppress the effect.)

 The expression in eq.(10) is not applicable for a GW pulse which does 
 not last for many cycles.  Here we recall that for Weber detector use 
 of materials with high quality factor $Q$ was important even for a short 
 GW pulse. For a sustained signal, in principle, one could get enormous 
 amplitude enhancement at resonance with $Q$ factor of about $10^6$. 
 However, even for a short pulse, there is ringing effect for Weber 
 detector operating at resonance, which was helpful in enhancing signal
 to noise ratio. Due to this ringing effect, the detector continues
 to vibrate in the resonant mode for some time even after the passing of
 the pulse through the detector (due to energy absorbed from the pulse
 in the resonant mode). For example, for a GW burst lasting a few ms, 
 the resonant bar can continue to ring for a long time, of order 
 10 min \cite{weber,ring}. This particular mode having a definite frequency
 allows for separation between the random noise and the signal. In the
 same way, if the pulsar continues to {\it ring} for some time after the 
 GW pulse has passed through it, then the radio pulses will continue to
 retain this {\it definite frequency} signal hidden within. Folding of many 
 pulses may be able to separate this {\it ringing} signal. If there is 
 any such significant gain at resonance then there may be a possibility 
 of these pulsar detectors competing with earth based detectors even for 
 sources in our galaxy.

We also mention that, in a previous work \cite{ourns2015}, some of us have 
proposed that
phase transition dynamics inside a pulsar core can lead to density fluctuation
changes which can be detected by measuring changes in the pulse profiles
(and may even account for certain glitches/anti glitches). Further, we argued
that rapidly evolving density fluctuations can lead to changes in the quadrupole
moment  providing new powerful GW sources, possibly at very high
frequencies of order MHz. Combining with
our present results, we can infer that phase transitions happening inside
any NS core (not necessarily a pulsar) may provide additional GW sources
at very high frequencies which may be detectable by our presently
proposed remote pulsar GW detectors. 
 
As globular clusters have a very rich population of neutron stars, they may 
be an ideal place to look for such events. In particular, collapsed core 
globular clusters can have inter-star distance as small as 1/10 ly in the 
core, with larger probability of BNS mergers, or black hole mergers, and 
at the same time, a possibility of small pulsar detector - GW source 
separation. We may make a very rough estimate for the expected rates
as follows.  Note that for eqn.(9), the separation between pulsar and
source was taken to be about 1 ly when resonance effects were not
incorporated. If the resonance effects allow the sensitivity to be
increased even by a factor of about hundred (with ringing of pulsar as we
mentioned like for Weber bar) then the distance can easily
be allowed to be of the size of a typical globular cluster, that is about 100
parsec. Most of the pulsars (more that 50\%) and BNS are inside these
globular clusters. With about 2000 pulsars in these globular clusters 
(which number about 200) in the Milky Way, one can expect that for any BNS 
merger in a globular cluster, there will be few pulsars there which will
be able to detect the emitted GWs (again, assuming some resonance effect). 
The expected rate of BNS merger in the Milky Way is about 1 per 10,000 years 
and may even be 1 per 3000 years \cite{abbott2017} and this estimate suggests
that most of these should be detectable by pulsars. This estimate does not 
look optimistic from the point of view of observations, especially
when we know that very few of these pulsars are stable millisecond
pulsars with the desired accuracy.  But a further increase in sensitivity
factor by few orders of magnitude, either by stronger resonance effect,
or by a higher precision for pulsar timing measurement, or simply
having a much stronger source, may make such pulsars in the milky way
sensitive to events in other nearby galaxies. That may bring the expected
rates of such pulsar-detections within observable reach. We mention
that Dyson had proposed long ago \cite{dyson} that by using an array
of seismometers, earth itself may be used as gravitational wave detector 
possibly for gravitational waves emitted by pulsars. More recently, proposals
have been made for the detection of gravitational waves by their effects on 
nearby stars, especially on the solar acoustic modes (helioseismology
and astroseismology) \cite{seism}. These proposals are very similar
in spirit to our proposals, though to our knowledge, the monitoring
of transient changes in the pulsar signals due to pulsar deformations
by gravitational waves (as proposed here) has not been made. 

 Let us estimate what level of sensitivity will be required for this
 pulsar-detection of GW if the source lies in Virgo cluster. With 
 about 2000 galaxies in the Virgo cluster, and assuming the rate
 of BNS merger in a typical galaxy ranging from 1 in 10,000 years to
 1 in 3000 years, one can expect one event in few years originating
 from Virgo cluster. With typical distance of source being about
 50-60 Mly, we can expect the strain amplitude to be $h \sim 10^{-18}
 - 10^{-19}$, similar to the one observed by the recent BNS merger.
With eqns.(8)-(9), we expect fractional change in moment of inertia,
and hence the fractional change in the spin rate to be about 
$10^{-20} - 10^{-21}$. This is too small to be detected by pulsar
timing measurements (assuming best accuracy for the measurement of fractional
change in pulse timing to be about 10$^{-13}$) unless there is a strong
effect of resonance. As we mentioned above, resonance is very likely as 
typical frequency of GW from BNS merger overlaps with the natural frequencies
of some of the modes of neutron star oscillations. Taking a clue from
the Weber bar detector, if the sensitivity can be enhanced by a factor
of $10^6$ or more then pulsars in Milky way can detect these signals.
An enhancement in the accuracy of measurement of pulsar timings 
will also help in this direction. Important thing to realize is that,
essentially all the pulsars in Milky way will oscillate with
the GW hitting them at widely different times. So the measurements
of pulsar timing fluctuations due to a single GW event can be 
carried out over period of many years. A systematic analysis of
just the arrival time of GW at different pulsars can help pin point
the location of the GW source. Of course additional information
about the source direction can be obtained from any wobbling induced 
in the pulsar rotation due to GW induced deformations.  Clearly, this method 
will be most useful if one could use pulsars very far away. As 
globular clusters seem very common in other galaxies, in that sense 
this technique should provide motivation for detailed and precision 
studies of pulsars very far away. We mention that for far away
pulsars measurements of quantities like timing residuals will not be
vary accurate due to lack of detailed knowledge about changes in the
intergalactic medium etc. (apart from the fact that one needs larger
telescopes just to detect weak signals from very far away pulsars).
However, pulsar detection method we have proposed uses changes in pulser 
timings over very short period, and changes in intergalactic medium etc. 
may not be much relevant for this.

 We conclude by pointing out the main features of our proposal that pulsars
spread out in space can be used as a family of (Weber) gravitational wave
detectors. GW is detected by its effect on the moment of inertia of the
NS, and resulting signal, in terms of variations in the pulsar pulse timing
and its profile (resulting from induced wobbling depending on the
source direction w.r.t the pulsar rotation axis) is transmitted to 
earth based pulsar observatories. The effect will be most pronounced at 
resonance.  It has the possibility of extending our reach 
further out in space for this new era of gravitational wave astronomy for 
the detection of GW. Even for relatively nearby GW sources, these pulsar
detectors, combined with other conventional GW detectors, can help
in improving the accuracy for determination of precise location of the 
GW source. This can be of crucial importance for sources which do
not emit any other form of radiation apart from GW, such as black hole 
mergers.  We also point out that our proposal provides an opportunity to, 
in some sense, revisit the  gravitational wave events which have already 
been detected by LIGO (and VIRGO). For this one should look for specific 
pulsars which would have been disturbed by these events, and will transmit 
this disturbance via their pulse signals in any foreseeable future. If these
future pulsar events can be predicted with accuracy then a focused effort 
can be made to detect any possible changes in the signals of those specific
pulsars. This will be a direct test of any feasibility of the present proposal.
Any possible such detection will give valuable information about
the neutron star parameters (such as equation of state).

\begin{acknowledgements}
We thank Kandaswamy Subramanian, Rajaram Nityananda, Balasubramanian Iyer,
Joseph Samuel, Raghunathan  Srianand, Raghavan Rangarajan and Bharat Kumar 
for very useful suggestions and comments.
\end{acknowledgements}
 

\begin{thebibliography}{99}
\bibitem{abbott2016}
B. P. Abbott et al. (LIGO Scientific Collaboration and Virgo Collaboration), 
Phys. Rev. Lett. {\bf 116}, 061102 (2016).
\bibitem{abbott2017}
B. P. Abbott et al. (LIGO Scientific Collaboration and Virgo Collaboration),
Phys. Rev. Lett. {\bf 119}, 161101 (2017).
\bibitem{weber}
J. Weber, Phys. Rev. Lett. {\bf 18}, 498 (1967); ibid Phys. Rev. Lett.{\bf 20}, 1307  (1968);
ibid Phys. Rev. Lett. {\bf 22}, 1320 (1969).
\bibitem{aguiar2011}
O. D. Aguiar, Res.Astron.Astrophys. {\bf 11}, 1 (2011);  arXiv:1009.1138
\bibitem{ring} M. Maggiore, {\it  Gravitational Waves: Volume 1: Theory 
and Experiments} Oxford publications, 2007;
DOI: 10.1093/acprof:oso/9780198570745.001.0001.
\bibitem{pta} G. Hobbs and S. Dai, arXiv:1707.01615
\bibitem{hinderer2008}
T. Hinderer, Astrophys.J. {\bf 677}, 1216-1220 (2008).
\bibitem{carrollbook}
S. M. Carroll, Spacetime and Geometry: An Introduction to General Relativity, Addison Wesley, 2004.
\bibitem{archibald2013}
R. F. Archibald et al., Nature {\bf 497}, 591 (2013).
\bibitem{tsang2012}
D. Tsang, J. S. Read, T. Hinderer, A. L. Piro, R. Bondarescu, Phys. Rev. Lett. {\bf 108}, 011102 (2012).
\bibitem{nstaroscl} D. Lai, Mon. Not. R. Astron. Soc.
{\bf 270}, 611 (1994).
\bibitem{ourns2015}
P. Bagchi, A. Das, B. Layek, A. M. Srivastava, Phys.Lett. {\bf B747}, 120-124 (2015).
\bibitem{dyson} F. Dyson, Astrophys. J. {\bf 156}, 529, 1969.

\bibitem{seism} L. Lopes and J. Silk, Astrophys. J. {\bf 794}, 32 (2014);
{\it ibid} {\bf 807}, 135 (2015); {\it ibid}, {\bf 844}, 39 (2017). 

\end{thebibliography}


\end{document}